\documentclass[aps,pre,subeqns.floatfix,amsmath,twocolumn]{revtex4}
\usepackage[pdftex]{graphicx}
\usepackage{amsmath}
\usepackage{amsfonts}
\usepackage{graphicx}
\usepackage{hyperref}
\usepackage{comment}
\usepackage{mathtools}
\usepackage{verbatim}
\usepackage[outdir=./]{epstopdf}
\usepackage{subfigure}
\usepackage{hyperref}
\usepackage{latexsym}
\usepackage{dcolumn}
\usepackage{epsf}
\usepackage{float}

\usepackage{color}
\usepackage[utf8]{inputenc}

\begin{document}
\title{Effect of phase-lag on synchronization in adaptive multilayer networks with higher-order interactions}
\author{Anath Bandhu Das}
\email{anathbandhu0498@gmail.com}
\author{Sangita Dutta}
\author{Pinaki Pal}
\email{ppal.maths@nitdgp.ac.in}
\affiliation{Department of Mathematics, National Institute of Technology, Durgapur~713209, India}

\begin{abstract}
We investigate the transition to synchronization in adaptive multilayer networks with higher-order interactions both analytically and numerically in the presence of phase frustration ($\beta$). The higher order topology consists of pairwise and triadic couplings. The analytical framework for the investigation is based on the Ott-Antonsen ansatz which leads to a convenient low-dimensional model. Extensive bifurcation analysis of the low-dimensional model and the numerical simulation of the full networks are performed to explore the paths to synchronization. The combined analysis shows a complex dependence of the transition to synchronization on adaptation exponents, coupling strengths, phase lag parameter, and multilayer configuration. Various types of transitions to synchronization, namely continuous, tiered, and explosive, are exhibited by the system in different regions of the parameter space. In all the cases, a satisfactory match between the low-dimensional model and the numerical simulation results is observed. The origin of different transitions to synchronization is clearly understood using the low-dimensional model. Exploration of a wide region of the parameter space suggests that the phase frustration parameter inhibits tired as well as explosive synchronization transitions for fixed triadic coupling strength ($K_2$). On the other hand, discontinuous transition is promoted by the phase frustration parameter for fixed pairwise coupling strength ($K_1$). Moreover, the exponent of the adaptation function with the pairwise coupling decreases the width of the hysteresis, despite the dominance of the higher-order coupling for fixed $\beta$ and $K_2$. While, the exponent of the function adapted with higher-order coupling shows the opposite effect, it promotes bistability in spite of dominance of pairwise coupling strength for fixed $\beta$, and $K_1$. 
\end{abstract}

\maketitle


\section{Introduction}
Over the past few decades, complex networks have become a focal point of scientific investigation due to their powerful ability to represent and analyze interconnected dynamical systems. This growing interest spans multiple disciplines such as physics, biology, ecology, social sciences, and engineering~\cite{strogatz2001exploring,boccaletti2006complex,wang2003complex}, where network-based approaches offer valuable insights into system behavior and structure. From a mathematical perspective, a network is typically represented as a graph in which the individual components of a system correspond to nodes, and the interactions or relationships between them are depicted as edges or links connecting these nodes. To enhance the conventional network framework, the notion of multilayer networks has been introduced by researchers. This generalized structure provides a more realistic model for numerous complex systems observed in the real world~\cite{boccaletti2014structure,bianconi2018multilayer}.  Representative examples include transportation infrastructures~\cite{cardillo2013modeling}, neural circuits in the brain~\cite{majhi2019chimera,rakshit2018synchronization,frolov2020revealing}, and various forms of social interaction networks~\cite{szell2010multirelational,lu2023multiplex}. A multilayer network is composed of multiple individual layers, each comprising its own nodes and intralayer connections, while also being linked to other layers via interlayer connections. The fundamental premise underlying this representation is that the intricate relationships both within and between layers can be effectively captured using pairwise interactions.
\par While the traditional multilayer network framework, composed of interlayer and intralayer pairwise interactions, has been instrumental in modeling a wide range of complex systems, including transportation, neuronal, and social networks, it still exhibits limitations when it comes to capturing more intricate forms of interaction. In many real-world systems, the assumption that connections can be fully described by pairwise links is insufficient~\cite{battiston2020networks,boccaletti2023structure}. Situations arise across diverse domains such as functional~\cite{petri2014homological,lee2012persistent} and structural brain networks~\cite{sizemore2018cliques}, protein-protein interactions~\cite{estrada2018centralities}, semantic networks~\cite{sizemore2018knowledge}, random walks~\cite{chitra2019random,carletti2020random,di2024dynamical}, collaboration networks~\cite{patania2017shape,vasilyeva2021multilayer}, epidemic spreading~\cite{iacopini2019simplicial,kim2024higher,wang2024epidemic}, and ecological communities~\cite{grilli2017higher,alvarez2021evolutionary}, where the interactions naturally occur among groups rather than just between pairs of entities. These examples underline the importance of moving beyond pairwise modeling toward a framework that can accommodate group-level interactions. Such complex interaction patterns are better represented through mathematical structures like simplicial complexes~\cite{jonsson2008simplicial,battiston2021physics,bianconi2021higher,zhao2021simplicial} and hypergraphs~\cite{berge1973graphs,ghoshal2009random,adhikari2023synchronization}. In these frameworks, group interactions are described using geometric units of various dimensions. Simplicial complexes use simplices, where a simplex of order $d$ includes exactly $(d+1)$ nodes and also contains all its lower-order subsets. Hypergraphs represent these relationships through hyperedges that can connect any number of nodes without requiring the presence of all pairwise links. This flexibility makes hypergraphs more general, while simplicial complexes impose a structured hierarchy. Both approaches provide powerful tools to study systems where interactions extend beyond the pairwise level.
\par Among the various collective phenomena studied in multilayer networks, synchronization~\cite{boccaletti2002synchronization,pikovsky2003universal,strogatz2004sync,boccaletti2018synchronization,wu2024synchronization} stands out as a particularly compelling subject that has attracted considerable scholarly interest. Synchronization is a phenomenon widely observed across both natural and engineered systems, including the coordinated flashing of fireflies~\cite{buck1988synchronous}, rhythmic clapping in auditoriums~\cite{neda2000sound}, neural activity in the brain~\cite{penn2016network,pal2021synchronization}, cellular oscillations in yeast populations~\cite{de2007dynamical}, as well as in mechanical systems like coupled metronomes and large-scale infrastructures such as power grids~\cite{motter2013spontaneous}. To investigate the mechanisms underlying such coherent dynamics in complex systems, numerous models have been developed based on nonlinear dynamical units interacting over networked structures. When uncoupled, these individual units typically exhibit uncoordinated or incoherent behavior. However, once interactions are introduced, the system can undergo a collective transformation, giving rise to synchronization through a phase transition. This transition reflects the system’s evolution from a disordered, asynchronous regime to a state of coherent collective motion. A foundational approach to capturing this transition was introduced by Kuramoto~\cite{kuramoto2003chemical}, who proposed a simplified phase oscillator model, where each unit operates with its own intrinsic frequency $\omega_i$ and interacts with others via a periodic coupling function that captures pairwise interactions. Depending on the structural configuration and dynamical parameters of the system, the route to synchronization from an initially asynchronous state in such models can be continuous, discontinuous, or even explosive~\cite{ichinomiya2004frequency,gomez2007paths,gomez2011explosive,rodrigues2016kuramoto}. 
\par Moreover, allowing the coupling strength to adapt dynamically makes it possible to model oscillator systems that exhibit positive feedback in their interactions~\cite{zhang2015explosive,manoranjani2023phase,biswas2024effect,chen2024synchronization}. For instance, the coupling strength can be made dependent on the number of oscillators participating in the synchronized cluster. When applied to systems with higher-order interactions, such adaptive schemes have been shown to depict tiered synchronization~\cite{rajwani2023tiered} and double explosive synchronization~\cite{dutta2025double}. Introducing a phase lag in pairwise interactions leads to phase frustration, which in turn shifts the critical coupling strength required for synchronization~\cite{omel2016there,sakaguchi1986soluble}. Furthermore, the adaptation of dynamical states has been investigated in systems where higher-order interactions are combined with phase lag, revealing rich and complex synchronization behaviors~\cite{dutta2023impact,sharma2024synchronization,dutta2024transition,li2025analysis}. Similar to the diverse synchronization transitions seen in adaptive dynamical systems, multilayer network architectures have also been shown to facilitate explosive synchronization. Such abrupt transitions to synchrony have been reported across a range of multilayer configurations, including master–slave arrangements~\cite{wu2022double}, inter-pinned layers~\cite{kachhvah2021explosive}, phase-frustrated networks~\cite{khanra2018explosive,khanra2021explosive} as well as in systems featuring intralayer adaptive coupling~\cite{zhang2015explosive} and interlayer adaptive coupling~\cite{kumar2020interlayer}. Recently, Ghosh \emph{et al.}~\cite{ghosh2025transitions} showed that incorporating order parameter-based adaptation in higher-order multilayer networks significantly alters synchronization dynamics and the routes to coherence. They investigated both linear and nonlinear adaptation forms, revealing tiered synchronization states arising from multistability and diverse bifurcation scenarios. Notably, different adaptation exponents led to unique combinations of continuous, discontinuous, and explosive transitions, influenced by the nature of 1- and 2-simplex interactions and the underlying adaptation strength.
\par In this paper, we have investigated the role of phase-lag on synchronization transitions by considering a adaptive multilayer network with higher-order interactions. For simplification we took upto triangular interactions the number of layers upto two. We have adapted the global order parameter with both the pairwise and triadic coupling strengths in form of power law functions. The cross adaptation technique of the order parameters offer the interaction between two layers. We have obtained a low dimensional model of the original $N$-dimensional system by using Ott-Antonsen ansatz~\cite{ott2008low}. The stability analysis of this reduced model easily explores the stable branches of order parameters. Numerically simulated data of the $N$-dimensional model nicely follow these stable branches, generating different continuous and discontinuous paths to synchronization like tiered and explosive transition depending on the choice of initial conditions. Both the analytical and numerical investigation reveal that the phase-lag has opposite effect on the synchronization transitions with the variation of pairwise and triadic coupling strength. The combined effect of adaptation exponents, coupling strengths, phase-lag parameters and the multilayer configuration of the networked system have been studied in detail. In the next section we have describe the model and go for deriving the low dimensional model.


  \section{Model}
  In this study, we explore the dynamics of an adaptive multilayer network, where each layer comprises globally coupled Sakaguchi-Kuramoto~\cite{kuramoto2003chemical,sakaguchi1986soluble} oscillators influenced by adaptive feedback. The interactions within the network include both pairwise (1-simplex) and three-body (2-simplex) coupling mechanisms. Building on the adaptive multilayer structure, the evolution of the network is governed by equations in which the connection weights adjust dynamically in response to the global order parameter. This feedback mechanism dynamically regulates the interaction strengths among the oscillators distributed across the $l$ layers of the network. The evolution of the oscillator phases is governed by the following set of equations:
  
  \begin{align}
    \dot{\theta}_{i,l}&=\omega_{i,l}+\frac{K_{1}f_{p,l}(\vec{r}(t))}{N}\sum_{j=1}^N \sin(\theta_{j,l}-\theta_{i,l}-\beta_{l}) \nonumber\\
     &+\frac{K_{2}f_{h,l}(\vec{r}(t))}{N} \sum_{j=1}^N \sum_{k=1}^N \sin(2\theta_{j,l}-\theta_{k,l}-\theta_{i,l}-\beta_{l}), \label{eqn1}
  \end{align}
  where $i=1,2,\dots,N$ and $l=1,2,\dots,L$. Here, $N$ denotes the number of oscillators in each layer, and $L$ is the total number of layers. Each oscillator $i$ in layer $l$ is characterized by its phase $\theta_{i,l}$, which evolves over time, and its intrinsic frequency $\omega_{i,l}$, typically drawn from a unimodal distribution $g(\omega)$. The coupling strengths for pairwise and higher-order interactions are denoted by $K_{1}$ and $K_{2}$, respectively. The phase lag parameter $\beta_{l}$ introduces asymmetry in the interactions within each layer. The functions $f_{p,l}(\vec{r}(t))$ and $f_{h,l}(\vec{r}(t))$ are the adaptive functions associated with the pairwise and higher-order interactions, respectively. These depend on the instantaneous global order parameter vector $\vec{r}(t)=\{r_{1,1},r_{1,2},\dots,r_{1,L} \}^{T}$, where each component $r_{1,l}$ is defined by,
  \begin{equation}
      z_{1,l}=r_{1,l}e^{\iota \psi_{1,l}}=\frac{1}{N} \sum_{j=1}^{N}e^{\iota \theta_{j,l}}. \label{eqn2}
  \end{equation}
  Here, $r_{1,l}$ and $\psi_{1,l}$ denote the modulus and phase angle of the complex order parameter $z_{1,l}$, respectively, with $l=1,2,\dots,L$. 
  $r_{1,l}=1$ signifies complete synchronization across all nodes, whereas $r_{1,l}=0$ reflects a state of total incoherence. 
  Moreover, $z_{2,l}=r_{2,l}e^{\iota \psi_{2,l}}=\frac{1}{N} \sum_{j=1}^{N}e^{2\iota \theta_{j,l}}$ captures the two-cluster synchronization states of the oscillators.
  \par It is worth emphasizing that in Eq.~(\ref{eqn1}), the interaction between layers is mediated solely through the adaptation functions $f_{p,l}$ (for pairwise interactions) and $f_{h,l}$ (for triadic interactions). There is no direct phase coupling between oscillators belonging to different layers; rather, the interlayer influence is incorporated indirectly through the layer-specific order parameters $r_{1,l}$.
  \par In our investigation, we consider a two-layer network where the adaptation functions take the form: $f_{p,1}(\vec{r})=(A+Br_{1,2})^{p_{1}}$, $f_{p,2}(\vec{r})=(A+Br_{1,1})^{p_{2}}$, $f_{h,1}(\vec{r})=(A+Br_{1,2})^{h_{1}}$ and $f_{h,2}(\vec{r})=(A+Br_{1,1})^{h_{2}}$, with $r_{1,1}$ and $r_{1,2}$ denoting the global synchronization order parameters for layers 1 and 2, respectively. Here, $A, B\in\mathbb{R^{+}}$ and $p_{1},p_{2},h_{1},h_{2}\in\mathbb{R}$.  
  This formulation offers a simplified yet physically insightful version of Eq.~(\ref{eqn1}). This adaptation scheme is referred to as \textit{cross-adaptation}~\cite{zhang2015explosive,khanra2018explosive,khanra2021explosive,biswas2024effect}. Real-world examples include cascading failures in interconnected power grids~\cite{buldyrev2010catastrophic} or synchronized cheering in competitive events~\cite{zumeta2016collective}, where increased coordination in one group drives stronger synchronization in the other through mutual feedback.  When $A=0$, $B=1$, the adaptation functions simplify to the well-known power-law form, which has been extensively studied and applied in various network dynamics contexts~\cite{filatrella2007generalized,zou2020dynamics,cai2022exact}.


  \section{Evolution Equation of the Macroscopic Order Parameters}
  In this section we aim to reduce the dimension of the system Eq.~(\ref{eqn1}) to study its dynamic behavior. For this purpose, let us first rewrite Eq.~(\ref{eqn1}) using the definition of the complex order parameters $z_{1,l}$ and $z_{2,l}$, yields
  \begin{equation}
   \dot{\theta}_{i,l}=\omega_{i,l}+\frac{1}{2\iota} \left[H_{l} e^{-\iota(\theta_{i,l}+\beta_{l})} - H_{l}^{*} e^{\iota(\theta_{i,l}+\beta_{l})} \right], \label{eqn3}
  \end{equation}
  where, $H_{l}=k_{1}z_{1,l}+k_{2}z_{2,l} z_{1,l}^{*}$ with $k_{1}=K_{1}f_{p,l}$ and $k_{2}=K_{2}f_{h,l}$. $H_{l}^{*}$ is the complex conjugate of $H_{l}$.
  In order to investigate the role of the phase-lag parameter on the emergent dynamics of the system Eq.~(\ref{eqn1}) we consider the mean-field approach, i.e we have studied the system dynamics in the continuum limit ($N\rightarrow \infty$). At first, we characterized the oscillator system by density function $f_l(\theta, \omega, t)$, representing the density of the oscillators with phase $\theta$, natural frequency $\omega$ at time $t$ of layer $l$, satisfying the normalization condition  
\begin{equation}
       \int_{0}^{2\pi} f_l(\theta,\omega,t) d\theta=g_l(\omega) 
   \end{equation}
 for any value of $\omega$ and $t$. Since the number of oscillators is conserved in the system, the density functions must satisfy the continuity equation given by 
 \begin{equation}
       \frac{\partial f_{l}}{\partial t}+\frac{\partial}{\partial \theta_{l}} (f_{l}v_{l})=0, \label{eqn5}
   \end{equation}
   where $v_{l}(\theta_{l},\omega_{l},t)=\frac{d\theta_{l}}{dt}$ denotes the angular velocity of an oscillator located at phase $\theta_{l}$ and natural frequency $\omega_{l}$ at time $t$.
   \par Now, to facilitate analytical treatment of the model, we assume the natural frequencies follow a Lorentzian distribution. Specifically, for each layer $l$, the distribution $g_{l}(\omega)$ is defined as:
   \begin{equation}
       g_{l}(\omega)=\frac{\Delta_{l}}{\pi[\Delta_{l}^2+(\omega_{l}-\omega_{0,l})^2]},~\Delta_{l}>0, \label{eqn6}
   \end{equation}
   where $\omega_{0,l}$ represents the peak and $\Delta_{l}$ denotes the half-width of the frequency distribution of the $l^{\mathrm{th}}$ layer. It is worth emphasizing that the frequency distribution $g_{l}(\omega)$ is symmetric about its central frequency $\omega_{0,l}$. This symmetric structure makes it well-suited for modeling oscillator populations where a single dominant frequency is present---a feature commonly encountered in various physical systems.
   \par Since, the density function $f_{l}$ is $2\pi$-periodic with respect to $\theta$, we have expanded it in a Fourier series, given by
   \begin{equation}
       f_{l}=\frac{g_{l}(\omega)}{2\pi} \left[1+\sum_{n=1}^\infty a_{n,l}e^{\iota n\theta_{l}}+\sum_{n=1}^\infty a_{n,l}^{*}e^{-\iota n\theta_{l}} \right], \label{eqn7}
   \end{equation}
   where $a_{n,l}$ is the $n^{\mathrm{th}}$ Fourier coefficient and $a_{n,l}^{*}$ denotes the complex conjugate of $a_{n,l}$. Next, following the famous Ott-Antonsen ansatz~\cite{ott2008low}, we have expressed the Fourier coefficients as 
   $a_{n,l}=\alpha_{l}^{n}$, where $\alpha_{l}$ decays geometrically satisfying the condition $\alpha_{l}\ll1$. This assumption on the coefficients assures the convergence of the above series.  
   Then, substituting the expression of $f_{l}$ from Eq.~(\ref{eqn7}) into the continuity equation given by Eq.~(\ref{eqn5}), we arrive at a reduced one-dimensional differential equation describing the dynamics of $\alpha_{l}$, expressed as,
   \begin{equation}
       \dot{\alpha_{l}}=-\iota \omega_{l}\alpha_{l}-\frac{1}{2}\left(\alpha_{l}^2e^{-\iota\beta_{l}}H_{l}-e^{\iota\beta_{l}}H_{l}^* \right). \label{eqn8}
   \end{equation}
  Since, the synchronization behavior of the system can be described by the order parameter values, let us recall the definition of it. In the continuum limit, the complex order parameters can be expressed by the integral
   \begin{align}
       z_{p,l}&=\int_{-\infty}^{\infty} \int_{0}^{2\pi} f_{l}(\theta,\omega,t)e^{\iota p\theta}d\theta d\omega \nonumber \\ 
       &=\int_{-\infty}^{\infty}{\alpha_{l}^*}^p g_{l}(\omega) d\omega,~p=1,2. \label{eqn9}
   \end{align}
   \par This integration can further be evaluated using Cauchy’s residue theorem by taking a closed contour in the lower half of the $\omega$-plane, which yields $z_{1,l}=\alpha_{l}^*(\omega_{0,l}-\iota\Delta_{l},t)$ and $z_{2,l}={\alpha_{l}^*}^2(\omega_{0,l}-\iota\Delta_{l},t)=z_{1,l}^2$.
    Now inserting $\omega_{l}=\omega_{0,l}-\iota\Delta_{l}$ into Eq.~(\ref{eqn8}) yields the evolution equation governing $z_{1,l}$ as,
   \begin{align}
       \dot{z}_{1,l}=\iota \omega_{0,l} z_{1,l}&-\Delta_{l}z_{1,l}+\frac{1}{2}[(k_{1}z_{1,l}+k_{2}z^{2}_{1,l} z_{1,l}^{*})e^{-\iota \beta_{l}} \nonumber \\
       &-z^{2}_{1,l} (k_{1}z_{1,l}^*+k_{2}{z_{1,l}^*}^2 z_{1,l})e^{\iota \beta_{l}}]. \label{eqn10}
   \end{align}
   Substituting the relations $z_{1,l}=r_{1,l}e^{\iota \psi_{1,l}}$, $k_{1}=K_{1}f_{p,l}$ and $k_{2}=K_{2}f_{h,l}$ into the above equation and separating the real and imaginary components, we obtain the following set of equations that describe the evolution of $r_{1,l}$ and $\psi_{1,l}$ as,
   \begin{align}
       \dot{r}_{1,l}&=-\Delta_{l} r_{1,l}+\frac{r_{1,l}(1-r^{2}_{1,l})\cos\beta_{l}}{2}\left[K_{1}f_{p,l}+K_{2}f_{h,l}r^{2}_{1,l} \right], \label{eqn11} \\
       \dot{\psi}_{1,l}&=\omega_{0,l}-\frac{1}{2}\left[(K_{1}+K_{2}r^{2}_{1,l})(1+r^{2}_{1,l})\sin\beta_{l} \right]. \label{eqn12}
   \end{align}
   The above system of two coupled nonlinear ordinary differential equations characterizes the behavior of model (\ref{eqn1}) in terms of its macroscopic variables. From this point, we denote $r_{1,l}$ as $r_{l}$ and consider a two-layer network ($L=2$).
   Under this setup, the general evolution equation for the order parameter $r_{l}$, as given by Eq.~(\ref{eqn11}), can now be explicitly written for both layer 1 and layer 2 as,
    Therefore, 
   \begin{equation}\label{eqn13}
   \scalebox{0.85}{$
   \begin{aligned}
       \dot{r}_{1} &= -\Delta_{1} r_{1} + \frac{r_{1}(1 - r^{2}_{1}) \cos\beta_{1}}{2} \bigl[ K_{1}(A + Br_{2})^{p_{1}} + K_{2}(A + Br_{2})^{h_{1}} r^{2}_{1} \bigr],\\
       \dot{r}_{2} &= -\Delta_{2} r_{2} + \frac{r_{2}(1 - r^{2}_{2}) \cos\beta_{2}}{2} \bigl[ K_{1}(A + Br_{1})^{p_{2}} + K_{2}(A + Br_{1})^{h_{2}} r^{2}_{2} \bigr].
   \end{aligned}
   $}
   \end{equation}
   \begin{figure*}
    \centering
    \includegraphics[width=1\textwidth]{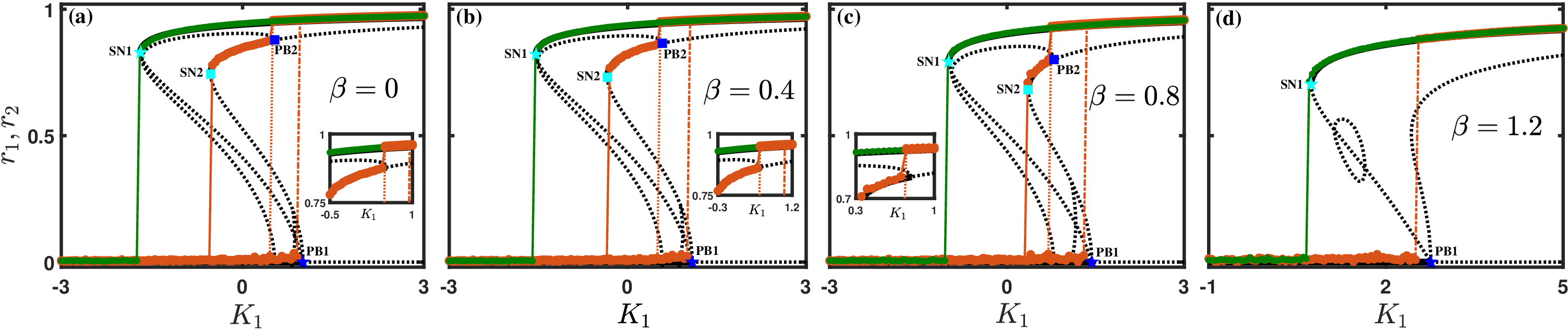}
    \caption{Synchronization profile with the variation of pairwise coupling $K_1$ for (a) $\beta=0$, (b) $\beta=0.4$, (c) $\beta=0.8$ and $\beta=1.2$. Other parameters are fixed at $p_{1}=p_{2}=1$, $h_{1}=h_{2}=1$, and $K_{2}=5$. Black solid (stable) and dotted (unstable) $r_1,~r_2$ curves are plotted by solving the reduced order model. Green and brown filled circles joined by solid, dotted and dashed dotted lines correspond to the numerically simulated order parameter values from system (\ref{eqn1}). Saddle-node (Pitchfork) bifurcation points are indicated by cyan (blue) circle and square. As $\beta$ value increases the weak synchronization state vanishes and the hysteresis width of the transition decreases. 
    }
    \label{fig1}
\end{figure*}
   Now we conduct the linear stability analysis of Eq.~(\ref{eqn13}). In the steady state, $\dot{r}_{1}=\dot{r}_{2}=0$ results in
   \begin{equation}\label{eqn14}
   \scalebox{0.93}{$
   \begin{aligned}
       G_{1}=\dot{r}_{1}=&-\Delta_{1} r_{1} + \frac{r_{1}(1 - r^{2}_{1}) \cos\beta_{1}}{2} \bigl[ K_{1}(A + Br_{2})^{p_{1}} \\
       &+ K_{2}(A + Br_{2})^{h_{1}} r^{2}_{1} \bigr]=0,\\
       G_{2}=\dot{r}_{2}=&-\Delta_{2} r_{2} + \frac{r_{2}(1 - r^{2}_{2}) \cos\beta_{2}}{2} \bigl[ K_{1}(A + Br_{1})^{p_{2}} \\
       &+ K_{2}(A + Br_{1})^{h_{2}} r^{2}_{2} \bigr]=0.
   \end{aligned}
   $}
   \end{equation}
   From Eq.~(\ref{eqn14}), it follows that $(r_{1},r_{2})=(0,0),(r^{*}_{1},0)$, and $(0,r^{*}_{2})$ constitute three trivial steady-state solutions of the two-dimensional reduced system (\ref{eqn13}), where $r^{*}_{l}$ is given by,
   \begin{equation}\label{eqn15}
   \scalebox{0.7}{$
   \begin{aligned}
       r^{*}_{l}=\sqrt{\frac{\bigl[ (K_{2}A^{h_{l}}-K_{1}A^{p_{l}})\cos\beta_l \bigr]\pm \sqrt{\cos^{2}\beta_l(K_{2}A^{h_{l}}+K_{1}A^{p_{l}})^2-8\Delta_lK_{2}A^{h_{l}}\cos\beta_l}}{2K_{2}A^{h_{l}}\cos\beta_{l}}},\\ l=1,2.
   \end{aligned}
   $}
   \end{equation}
   
   Beyond the aforementioned trivial steady states, one can also identify steady states of the form $(r^{*}_{1},r^{*}_{2})\neq(0,0)$. These non-trivial steady states, which depend on all system parameters, are generally not amenable to analytical solutions. Consequently, we determine them by numerically solving the coupled equations $G_{1}=0$ and $G_{2}=0$, as given in Eq.~(\ref{eqn14}).
   \par Now, to examine the stability of all steady-state solutions, we compute the Jacobian matrix associated with the dynamical system described in Eq.~(\ref{eqn13}) and check the eigen values of it in each case of the steady states. The elements of the Jacobian are given by the partial derivatives of $G_{l}$ with respect to $r_{j}$, where
   \begin{equation}\label{eqn16}
   \scalebox{0.75}{$
   \begin{aligned}
       \frac{\partial G_1}{\partial r_1}&=-\Delta_1+\frac{\cos\beta_1}{2}\bigl[ K_{1}(A+Br_{2})^{p_{1}}(1-3r^2_{1})+K_2(A+Br_{2})^{h_{1}}(3r^2_{1}-5r^4_{1}) \bigr], \\
       \frac{\partial G_1}{\partial r_2}&=\frac{r_1(1-r^2_{1})\cos\beta_{1}}{2}\bigl[ K_1 p_1 B(A+Br_2)^{p_1-1}+K_2 h_1 B(A+Br_2)^{h_1-1} \bigr], \\
       \frac{\partial G_2}{\partial r_1}&=\frac{r_2(1-r^2_{2})\cos\beta_{2}}{2}\bigl[ K_1 p_2 B(A+Br_1)^{p_2-1}+K_2 h_2 B(A+Br_1)^{h_2-1} \bigr], \\
       \frac{\partial G_2}{\partial r_2}&=-\Delta_2+\frac{\cos\beta_2}{2}\bigl[ K_{1}(A+Br_{1})^{p_{2}}(1-3r^2_{2})+K_2(A+Br_{1})^{h_{2}}(3r^2_{2}-5r^4_{2}) \bigr].
   \end{aligned}
   $}
   \end{equation}
   For the trivial steady state $(0,0)$,  the Jacobian matrix becomes diagonal, and hence, its eigenvalues are given directly by the diagonal elements, $\lambda_l=-\Delta_l+\frac{K_1A^{p_l}\cos\beta_l}{2}$ for $l=1,2$. The incoherent state remains stable as long as both eigenvalues are negative. However, when one of the eigenvalues becomes non-negative for certain parameter values, the incoherent state loses stability and becomes a saddle point. The corresponding parameter value then marks the onset of synchronization. Hence, the incoherent state $(0,0)$ remains stable as long as
   \begin{equation}
       K_1<\mathrm{min}\left\{\frac{2\Delta_l}{A^{p_l}\cos\beta_l}\right\},~\mathrm{for}~l=1,2. \label{eqn17}
   \end{equation}
   Otherwise, the state becomes unstable. The forward critical coupling threshold, denoted by $K_1^*=\mathrm{min}\left\{\frac{2\Delta_l}{A^{p_l}\cos\beta_l}\right\}$, is independent of the influence of higher-order interactions and their associated adaptation mechanisms. Furthermore, the stability properties of the other non-trivial steady states are determined numerically using the conditions derived from Eq.~(\ref{eqn16}).


\section{Results}
We have studied the synchronization behavior of the considered oscillator system both theoretically and numerically. To pursue the theoretical analysis, we have performed a comprehensive bifurcation analysis of the reduced-order model (\ref{eqn13}) using the \textsc{MATCONT} software~\cite{dhooge2003matcont}. Then to validate the theoretical findings, we have done extensive numerical simulation by integrating the system described by Eq.~(\ref{eqn1}) using the Euler method considering $N=10000$ oscillators per layer. The integration is performed over a total simulation time of $160000$ units with a time step of $dt=0.001$. The intrinsic frequencies of the oscillators are sampled from a Lorentzian distribution centered at zero ($\omega_0=0$) with half width $\Delta = 1$. For simplification, here, we assume symmetry in both the adaptation and phase-lag parameters, that is, $p_{1}=p_{2}$, $h_1=h_2$, and $\beta_1=\beta_2$. As a consequence, both layers exhibit identical dynamics; that is, $r_{1}$ and $r_{2}$ evolve in the same manner over time.

\begin{figure*}
    \centering
    \includegraphics[scale=0.35]{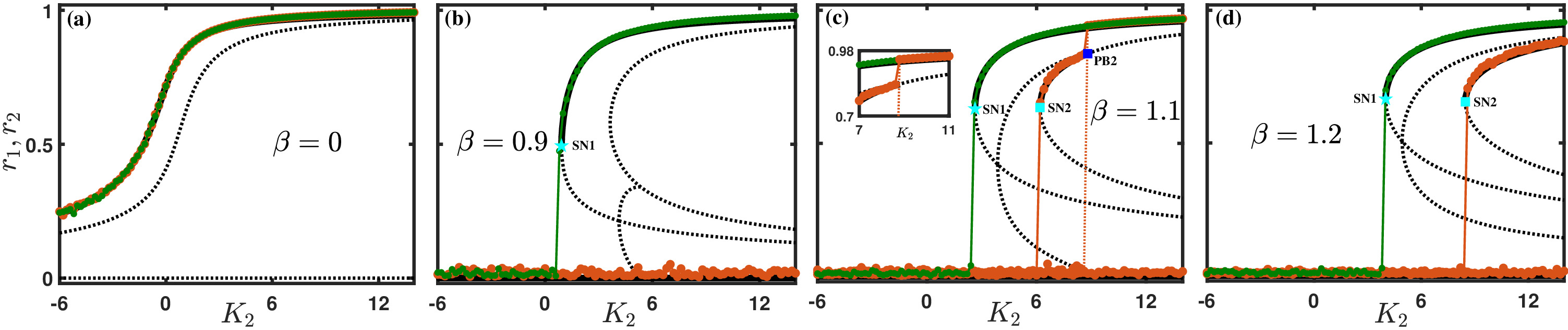}
    \caption{Synchronization profile with the variation of higher-order coupling $K_2$ for (a) $\beta=0$, (b) $\beta=0.4$, (c) $\beta=0.8$ and $\beta=1.2$. Other parameters are fixed at $p_{1}=p_{2}=1$, $h_{1}=h_{2}=1$, and $K_{1}=1.2$. Black solid (stable) and dotted (unstable) $r_1,~r_2$ curves are plotted by solving the reduced order model. Green and brown filled circles joined by solid, dotted and dashed dotted lines correspond to the numerically simulated order parameter values from system (\ref{eqn1}). Saddle-node (Pitchfork) bifurcation points are indicated by cyan (blue) circle and square. As $\beta$ value increases a weak synchronization state appears and the transition becomes explosive.} 
    \label{fig2}
\end{figure*}

\par At first we have theoretically explored the influence of the phase-lag parameter $\beta$ on the emergence of synchronization with the variation of the pairwise coupling $K_1$. In Fig.~\ref{fig1}, we present four cases: $\beta=0$, $\beta=0.4$, $\beta=0.8$, and $\beta=1.2$, where we have fixed the adaptation exponents at $p=1$, $h=1$ and the coupling strength $K_2$ at $5$ and trace the bifurcations with the vaiation of $K_1$. We observed several branches of the order parameters $r_1$ and $r_2$ are generated due to the cross adaptation of them in each layer. The stability of the branches have been examined directly by the MATCONT software. In all the figures of this paper the stable (unstable) branches have been indicated by solid (dotted) lines. From bifurcation point of view it is clear from Fig.~\ref{fig1} that the incoherent state and the coherent state are connected by multiple number of unstable branches depend on $\beta$ values. Moreover, the equilibrium points change their stability through saddle-node and pitchfork bifurcations. In particular, for $\beta=0$ the trivial equilibrium point $(r_1,r_2)=(0,0)$ loses it's stability via a subcritical pitchfork bifurcation (PB1) at $K_1=1$, indicated by a blue star in Fig.~\ref{fig1}(a). From this point, two unstable branches emerge in the backward direction. One of these branches corresponds to the symmetric non-trivial state $(r_1^*,r_2^*)$ which undergoes a saddle-node bifurcation (SN1) at $K_1=-1.6881$ (cyan star) and from there it continues as a stable strongly coherent branch. On the other hand, the second unstable branch generated from the PB1 point is associated with the asymmetric state $(r_1^*,0)$, which stabilizes through a saddle-node bifurcation (SN2) at $K_1=-0.5279$, shown by a cyan square. Again, it go though a subcritical pitchfork bifurcation (PB2) at $K_1=0.5322$ (blue square), beyond which it exists as unstable one. Therefore, comparatively a weak synchronized state appears in the range $K_1\in(-0.5279,0.5322)$. Due to symmetry, a similar behavior is observed for the branch $(0,r_2^*)$. Next we have increased the $\beta$ value to $0.4$ in Fig.~\ref{fig1}(b). A qualitatively similar bifurcation scenario is observed as in $\beta=0$. However, all the bifurcation points shifts to higher $K_1$ value, such as the pitchfork bifurcations PB1 and PB2 has shifted to $K_1=1.0857$ and $K_1=0.5820$, respectively. Also, the saddle-node bifurcation points SN1 and SN2 has shifted to $K_1=-1.5411$ and $K_1=-0.3401$, respectively. Further increase in $\beta(=0.8)$ (Fig.~\ref{fig1}(c)), shifted the whole synchronization diagram in the right hand side along with shrinking the range of $K_1$ corresponding to weak synchronization state. Finally, at $\beta=1.2$, only two unstable branches persists along with one stable state, which undergoes one saddle-node bifurcation (SN1) and one pitchfork bifurcation (PB1).   

\begin{figure*}
    \centering
    \includegraphics[scale=0.35]{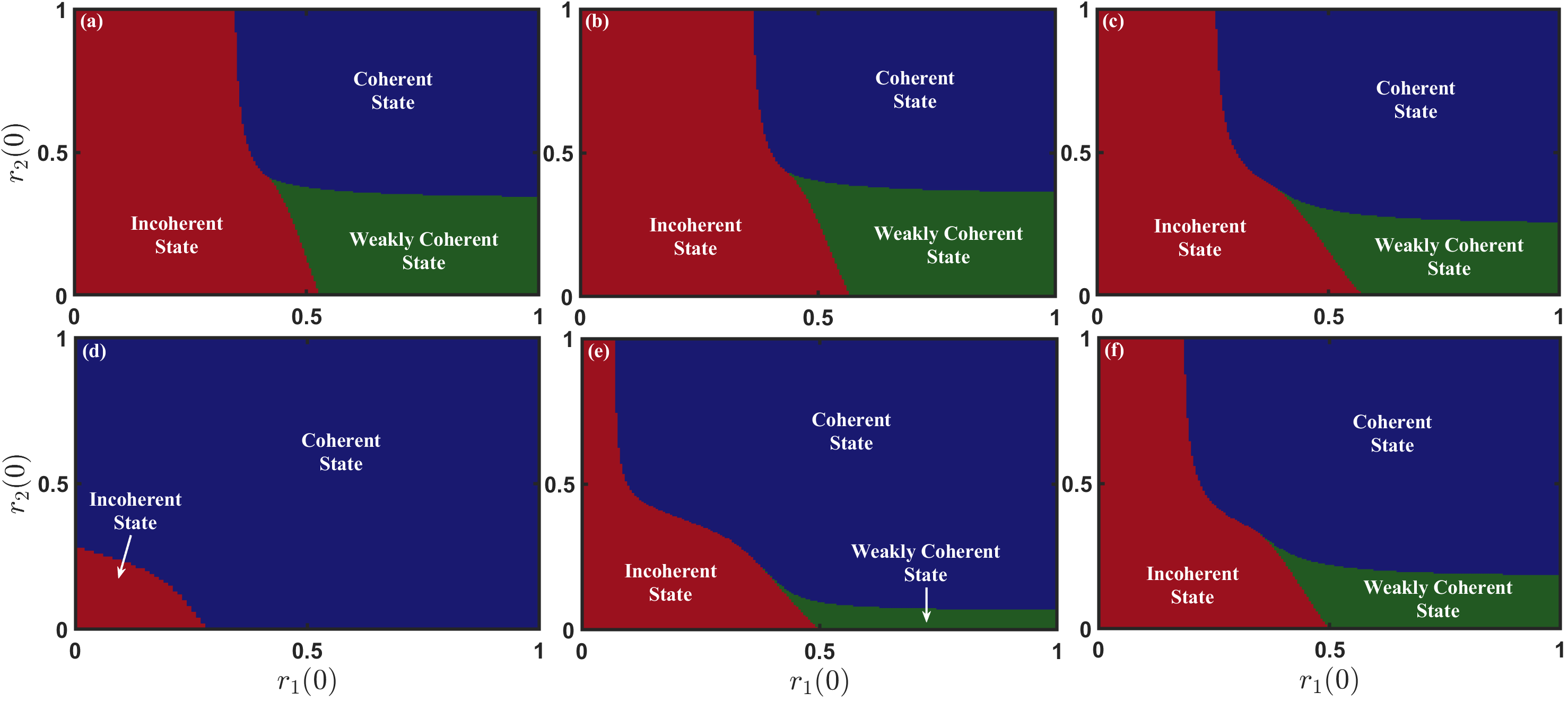}
    \caption{Basins of attraction for several stable synchronized states at different values of $\beta$. The red region represents the initial conditions ($r_1(0)$, $r_2(0)$) leading to a stable incoherent state, the blue region corresponds to the coherent state, and the green region indicates the weakly coherent state. Basin of attraction observed at (a) $K_1=0~\mathrm{and}~K_2=5$ for $\beta=0$, (b) $K_1=0~\mathrm{and}~K_2=5$ for $\beta=0.4$, (c) $K_1=0.5~\mathrm{and}~K_2=5$ for $\beta=0.8$ (d) $K_1=1.2~\mathrm{and}~K_2=7$ for $\beta=0.9$, (e) $K_1=1.2~\mathrm{and}~K_2=7$ for $\beta=1.1$ and (f) $K_1=1.2~\mathrm{and}~K_2=10$ for $\beta=1.2$.} 
    \label{fig3}
\end{figure*}

Now to see the synchronization transitions of the considered system, which contains finite number of oscillators, we performed numerical simulation once in the forward direction and then in the backward direction. In the forward direction, the simulation starts from $K_1=0$ and calculates the values of the order parameters $r_1$ and $r_2$ by increasing the coupling strength in a small increment up to $K_1=3$, where the system reaches to the strong synchronization state. Then in the backward direction, starting from $K_1=3$ we decrease the coupling value with the same step size as in the forward simulation and continue until $K_1=0$. Initially, the phases are distributed uniformly in a circle of unit radius. After that, at each coupling strength, the last phase value of the previous coupling has been used as the initial condition. First, we simulate the system by taking $\beta=0$ and superimpose the numerically calculated data points on the bifurcation diagrams. Other parameters are kept fixed at the same values as in Fig.~\ref{fig1}. We observed that with proper choice of initial condition the system transits abruptly from the incoherent state to weak or strong synchronized state. In case of abrupt jump from incoherent to weak synchronization state in the forward direction, the system follows the stable branch between two bifurcation points SN2 and PB2. Due to losing stability of this branch from the PB2 point, the system jumps to strong synchronization state, indicated by brown filled circles joined with solid line in Fig.~\ref{fig1}(a). In the backward direction, it follows the strong synchronization path upto SN1 point due to adiabatic continuation of the solutions. From there it jumps directly into the incoherent state (green-filled circles joined by a solid line), which takes the form of a tiered synchronization transition~\cite{dutta2024transition}. Therefore, in the range $0.5279\leq K1 \leq0.5322$, the system displays multistability, characterized by the coexistence of incoherent, weakly coherent, and strongly coherent states. In addition to the tiered transition, the system also follows two more explosive routes to synchronization depending on the choice of initial conditions, which are plotted as brown-filled circles joined by dashed and dashed dot lines. We follow the same simulation procedure to calculate the values of the order parameter for $\beta=0.4,~0.8$ and $1.2$. In all cases,, the numerical data points show an excellent agreement with the analytical curves predicted by the reduced-order analysis, confirming the reliability of the theoretical framework. It is clear from Fig.~\ref{fig1} that the increase in phase lag value decreases the hysteresis width along with shrinking the range of existence of weak synchronization state. After a certain $\beta$ value, the weak synchronization state corresponding to tiered transition vanishes and the system shows only explosive synchronization transition.

At this point, let us recall one result from~\cite{dutta2024transition} which states that the phase lag parameter has opposite effects on synchronization transitions with the variation of the coupling strengths $K_1$ and $K_2$. Following this result here also, we have investigated the effect of $\beta$ with the variation of $K_2$. We have done detailed bifurcation analysis using the reduced order model Eq.~(\ref{eqn13}) and presented it in Fig.~\ref{fig2}, for $\beta=0,~0.9,~1.1$, and $1.2$. This time we have fixed the value of $K_1$ at $1.2$. It is clear from the figure that for $\beta=0$, the reduced model has only two branches of equilibrium points, one of them is stable and another one is unstable. Numerical simulation in forward and backward direction reveals the continuous path to synchronization. The trivial steady state $(r_1,r_2)=(0,0)$ remains unstable throughout $K_2$ in this case. However, with an increase in the value of $\beta$, the trivial steady state becomes stable after the critical point $\beta=0.5857$ (approximately) along with appearance of multiple non zero branches in the diagram. For $\beta=0.9$, the nontrivial symmetric branch $(r_1^*,r_2^*)$ undergoes a saddle-node bifurcation (SN1) at $K_2=0.9195$, beyond which it persists as a stable coherent branch for higher values of $K_2$ and another two branches remain unstable in the range of their existence. Numerically calculated points reveal the type of transition as explosive, shown in Fig.~\ref{fig2}(b). Further increase in $\beta (=1.1)$ shifts the SN1 point at $K_2=2.6269$ (cyan star) corresponding to the nontrivial branch $(r_1^*,r_2^*)$ . Interestingly, the asymmetric branch $(r_1^*,0)$ becomes stable through a saddle-node bifurcation (SN2) at $K_2=6.1856$ (cyan square) and  persists until a subcritical pitchfork bifurcation (PB2) occurs at $K_2=8.8031$ (blue square), beyond which the branch loses its stability. Consequently, a weakly synchronized state, characterized by partial coherence, is observed within the interval $K_2\in(6.1856,8.8031)$ along with an explosive transition, clearly depicted from Fig.~\ref{fig2}(c). Also proper choice of initial condition the system shows additional explosive route to transition. Due to the system’s symmetry, a similar behavior is found for the asymmetric branch $(0,r_2^*)$. Moreover, this weakly synchronized state persists in comparatively large range of $K_2$ and eventually it tend to the strong synchronization state for high values of $\beta$ (Fig.~\ref{fig2}(d)).    

\begin{figure}
    \centering
    \includegraphics[scale=0.55]{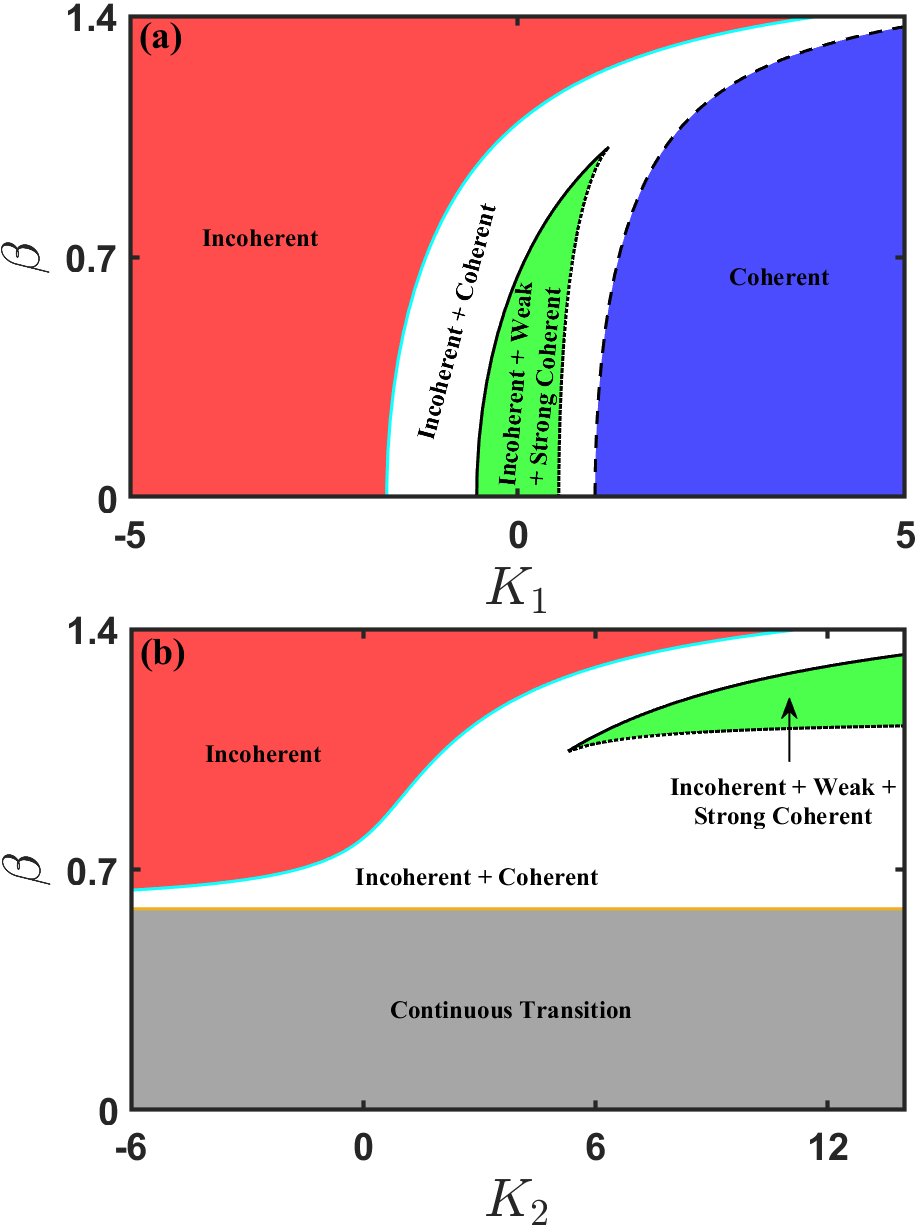}
    \caption{Stability diagram in (a) $K_1-\beta$ and (b) $K_2-\beta$ parameter space plotted using the reduced-order model. All other parameters are kept same as Fig.~\ref{fig1} and Fig.~\ref{fig2}. Solid cyan and solid black curves represent the movement of two saddle-node bifurcation points SN1 and SN2 respectively. Dotted black and dashed black curves respectively represent two pitchfork bifurcation points PB2 and PB1. Solid yellow curve represents the critical $\beta$ value for discontinuous transition. The bifurcation curves separates the whole spaces in four regimes based on the synchronization states, marked on the diagram.} 
    \label{fig4}
\end{figure}

Therefore, we have observed from Fig.~\ref{fig1} and Fig.~\ref{fig2} that the paths of synchronization transition in the multistability regime very much depend on the initial conditions. To properly understand this dependency, we have numerically integrated the reduced order model Eq.~(\ref{eqn13}) to obtain the basins of attraction and presented in Fig.~\ref{fig3}. We have plotted the values of $r_1$ for the basin. The axes represent the initial values $r_1(0)$ and $r_2(0)$, which serve as the initial conditions for the order parameters $r_1$ and $r_2$, respectively. The basin of initial conditions corresponding to the weakly coherent, strongly coherent and incoherent branch are shown by green, blue, and red regimes, respectively. It is worth noting that if one were to plot $r_2$ instead of $r_1$, a mirror symmetry of the basin across the main diagonal would be observed (not shown in the figure). In this context, the weakly coherent states illustrated in Fig.~\ref{fig3} represent steady states of the form $(r_1^*,0)$, while those involving $r_2$ would correspond to the steady states $(0,r_2^*)$. The first row of Fig.~\ref{fig3} is corresponding to Fig.~\ref{fig1}. We have taken three values from the $K_1-\beta$ space, namely $(K_1-\beta)=(0,0)$, $(0,0.4)$, and $(0.5,0.8)$, where the system exhibits multistability among three distinct states: a stable incoherent state, a weakly coherent state, and a strongly coherent state. Note that, low (high) values of $r_1^{(0)}$ and $r_2^{(0)}$ leads the system to the incoherent (synchronized) state. whereas, high $r_1^{(0)}$ along with low $r_2^{(0)}$ leads the system to weakly synchronized state, which implies that one layer must be initialized in a strongly synchronized state while the other begins in a incoherent state to access this branch. The second row of Fig.~\ref{fig3} is corresponding to Fig.~\ref{fig2}. It presents the initial values of the order parameters for three set of $(K_2-\beta)$ values $(7,0.9$, $(7,1.1)$ and $(10,1.2)$, which leads the system to three different states. Clearly, as $\beta$ increases, the regime of initial points for the incoherent state and the weak synchronized state increases. Therefore, Fig.~\ref{fig3} provides a complete set of initial conditions to reach at desired stable state in the synchronization diagram.

\begin{figure}
    \centering
    \includegraphics[width=0.5\textwidth]{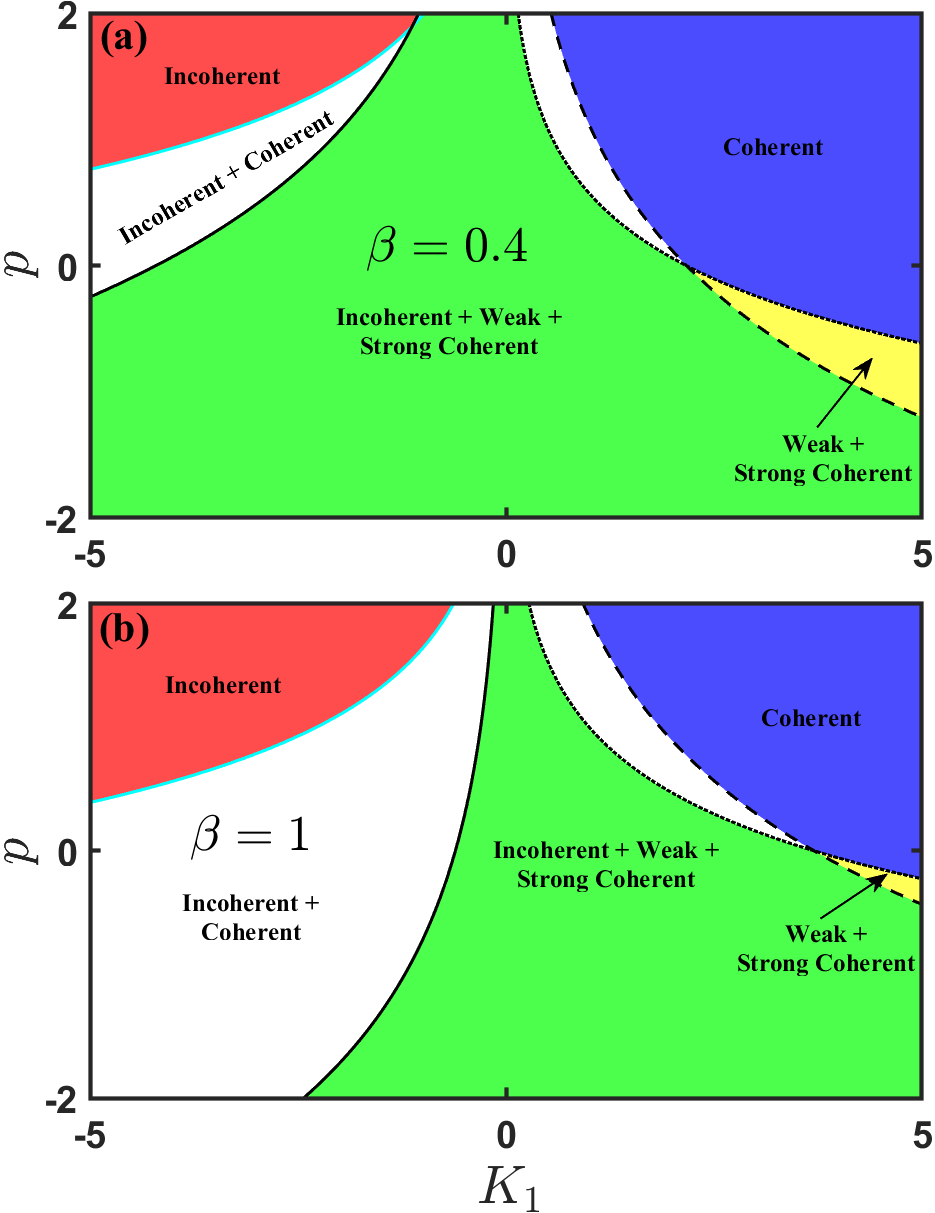}
    \caption{Stability diagram in $K_1-p$ parameter space for (a) $\beta=0.4$ and (b) $\beta=1$ plotted using the reduced-order model. All other parameters are fixed at $K_{2}=8$, $h=1$. Solid cyan and solid black curves represent the movement of two saddle-node bifurcation points SN1 and SN2 respectively. Dotted black and dashed black curves respectively represent two pitchfork bifurcation points PB2 and PB1. The bifurcation curves separates the whole spaces in five regimes based on the synchronization states, marked on the diagram.}
    \label{fig5}
\end{figure}

\begin{figure*}
    \centering
    \includegraphics[width=1\textwidth]{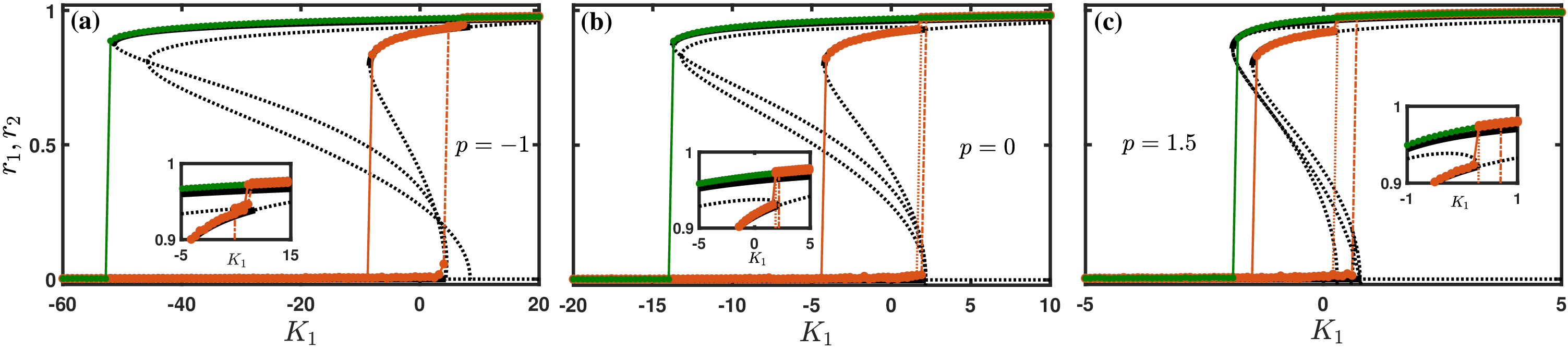}
    \caption{Synchronization profile with the variation of pairwise coupling $K_1$ for (a) $p=-1$, (b) $p=0$ and (c) $p=1.5$. Other parameters are fixed at $K_{2}=8$, $h=1$ and $\beta=0.4$. Black solid (stable) and dotted (unstable) $r_1,~r_2$ curves are plotted by solving the reduced order model. Green and brown filled circles joined by solid, dotted and dashed dotted lines correspond to the numerically simulated order parameter values from system (\ref{eqn1}). The fixed points changes stability through two saddle-node and pitchfork bifurcations. As $p$ value increases the hysteresis width of the transition decreases.}
    \label{fig6}
\end{figure*}

\begin{figure*}
    \centering
    \includegraphics[width=1\textwidth]{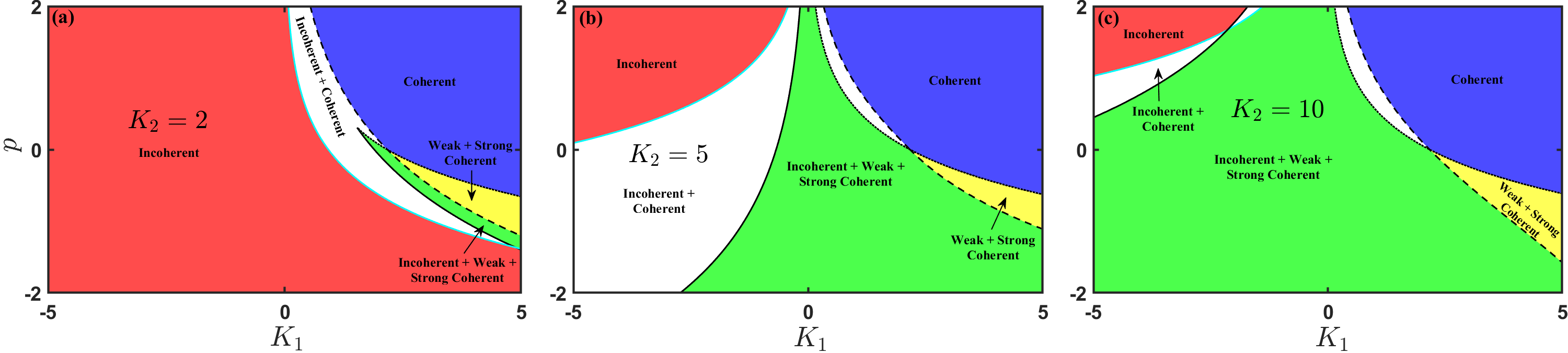}
    \caption{Stability diagram in $K_1-p$ parameter space for (a) $K_2=2$, (b) $K_2=5$ and (c) $K_2=10$ plotted using the reduced-order model. All other parameters are fixed at $\beta=0.4$, $h=1$. Solid cyan and solid black curves represent the movement of two saddle-node bifurcation points SN1 and SN2 respectively. Dotted black and dashed black curves respectively represent two pitchfork bifurcation points PB2 and PB1. The bifurcation curves separates the whole spaces in five regimes based on the synchronization states, marked on the diagram. This figure reflects that $K_2$ promotes multistability in $K_1-p$ parameter space.} 
    \label{fig7}
\end{figure*}
Now to get a better insight of Fig.~\ref{fig1} and Fig.~\ref{fig2}, we have prepared two parameter diagrams (Fig.~\ref{fig4}) using the reduced order model Eq.~(\ref{eqn13}). We have traced the saddle-node and pitchfork bifurcation points in respective parameter spaces. The solid cyan and solid black curve indicates the SN1 and SN2 points, respectively. The dotted black and dashed black curve respectively, denotes the PB2 and PB1 points. Fig.~\ref{fig4} depicts that these four bifurcation curves separated the $K_1-\beta$ and $K_2-\beta$ space into mainly four regimes based on the synchronization states. The red and blue regime indicates the incoherent and coherent state, respectively. In white regime the system displays bistability between incoherent and coherent state, while in the green regime the system shows multistability between incoherent, weak and strong synchronization state. In Fig.~\ref{fig4}(a) we observe that at $\beta=0$ the system experiences all the states. However, with an increase in $\beta$, the SN2 and PB2 curves come towards and meet at a critical point, vanishing the weak synchronization state from the synchronization diagram. Thereafter, the system only shows incoherent, coherent and bistability between these two states. Also, the regime bounded by the SN1 and PB1 curves shrinks in high $\beta$, implying a decrease in the width of hysteresis in the transition path. This analysis nicely explores the role of $\beta$ in the synchronization transitions with the variation of $K_1$ and the agreement of analytical and numerical points presented in Fig.~\ref{fig1} reconfirms it. Next, we move to Fig.~\ref{fig4}(b) and observe the opposite scenario. Here, the bifurcation curves separated the regimes mainly in two parts, one is associated with continuous transitions and another is associated with discontinuous transition. Again, the regime of discontinuous transition is separated in three parts. An increase in $\beta$ induces multistability in the system by broadening the regime between the bifurcation curves. The numerical simulation of Eq.~(\ref{eqn1}) again reconfirms this behavior of $\beta$ in Fig.~\ref{fig2}.  

Next, we turn our attention to understanding the combined role of the parameters $K_1,~K_2,~\beta,~p$ and $h$. For this we have prepared several two-parameter diagrams Fig.~\ref{fig5}, Fig.~\ref{fig7}, Fig.~\ref{fig11} and Fig.~\ref{fig10}. Besides that, we have validated this analysis by extensive numerical simulations. At first, let us check the effect of the exponent $p$ and phase-lag $\beta$ on synchronization transitions when $K_1$ is varied. Fig.~\ref{fig5} reports the movement of the bifurcation points in the $K_1-p$ space for $K_2=8,~h=1$, $\beta=0.4$ and $1$. Similarly to Fig.~\ref{fig4} here we have also marked the synchronization states in different regimes (five). For $\beta=0.4$, the system experiences multistability (green regime) in a wide regime at negative $p$. This green regime is getting narrower as the curves of the bifurcation points (SN1, SN2, PB1 and PB2) come closer with increasing $p$. Furthermore, all these curves moves to higher pairwise coupling value, where the SN1, SN2 curves intersect at $p=1.9092$ and the PB1, PB2 curves intersect at $p=0$. Therefore, the exponent $p$ decreases the width of the hysteresis and tends the transition to continuous. It is worth noticing that increase in $p$ reduces the effective pairwise coupling strength. Despite that the hysteresis width decreases due to multiplex nature of the networked system.  In addition to that, Fig.~\ref{fig5}(b) shows the significant effect of $\beta$ on the bifurcation points, i.e., the critical points of synchronization and desynchronization transitions. The whole scenario in the $K_1-p$ space has shifted to high $K_1$. As a result in the specified range of $K_1$ ($[-5,5]$), the regime of incoherent and bistable state (between incoherent and coherent) has expanded and other three regimes shrinks. This illustrates that in high $\beta$ the system requires higher coupling strength $K_1$ to synchronize (desynchronize) compared to lower $\beta$. For numerical verification we take three $p$ value such as $p=-1,~0$ and $1.5$ from Fig.~\ref{fig5}(a) and plot the synchronization diagrams in Fig.~\ref{fig6}. The numerical results match well with the analytical ones.

\begin{figure*}
    \centering
    \includegraphics[height=4.6cm,width=1\textwidth]{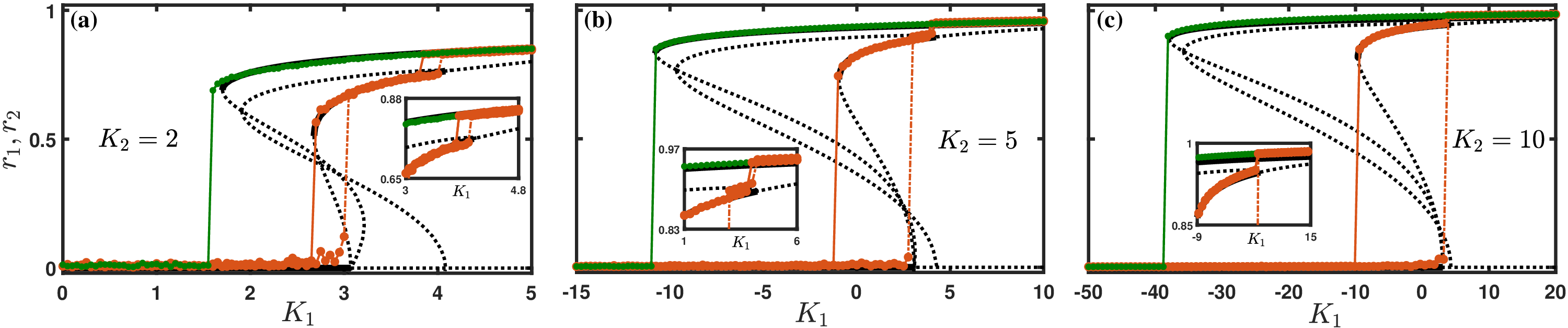}
    \caption{Synchronization profile with the variation of pairwise coupling $K_1$ for (a) $K_2=2$, (b) $K_2=5$ and (c) $K_2=10$. Other parameters are fixed at $p=-0.5$, $h=1$, and $\beta=0.4$. Black solid (stable) and dotted (unstable) $r_1,~r_2$ curves are plotted by solving the reduced order model. Green and brown filled circles joined by solid, dotted and dashed dotted lines correspond to the numerically simulated order parameter values from system (\ref{eqn1}). The fixed points changes stability through saddle-node and pitchfork bifurcations. As $K_2$ value increases the hysteresis width of the transition increases.}  
    \label{fig8}
\end{figure*}

\begin{figure}
    \centering
    \includegraphics[width=0.5\textwidth]{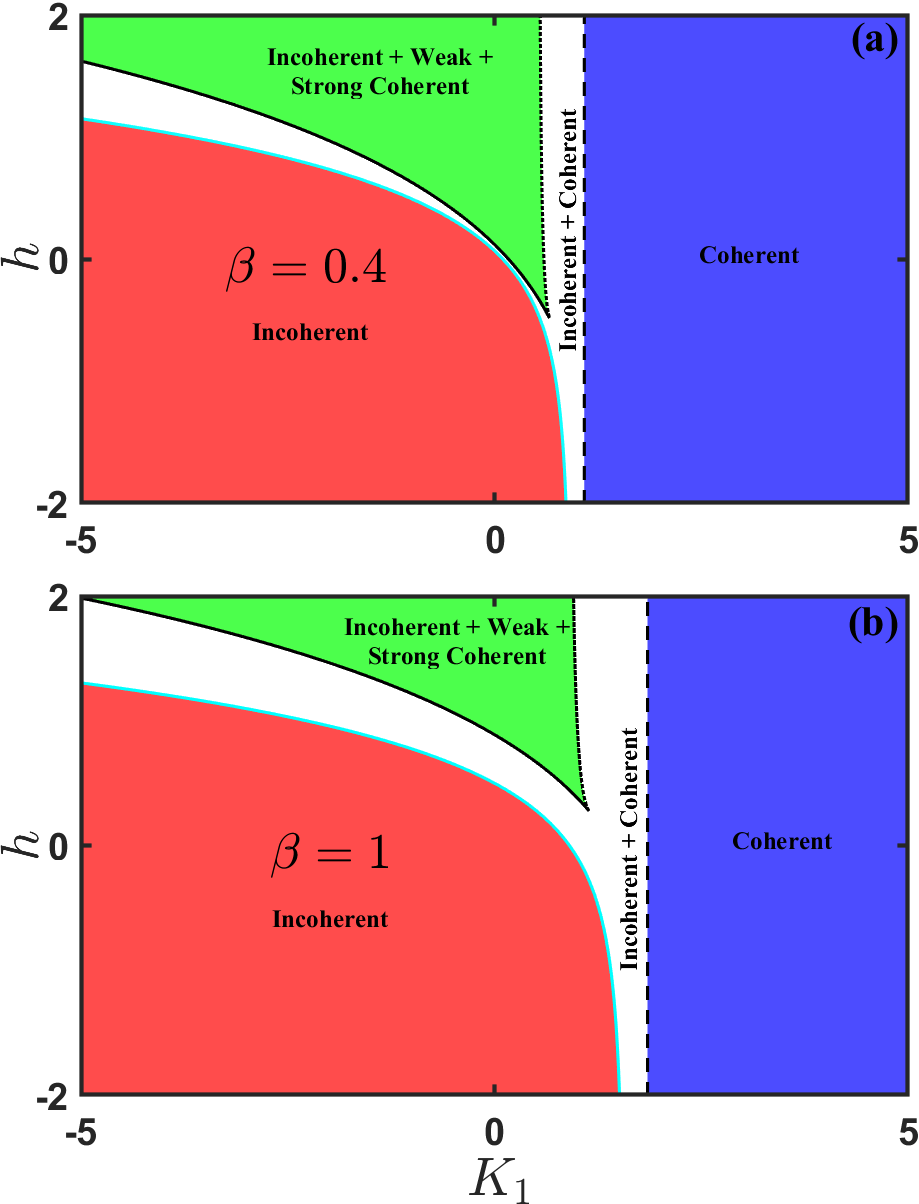}
    \caption{Stability diagram in $K_1-h$ parameter space for (a) $\beta=0.4$ and (b) $\beta=1$ plotted using the reduced-order model. All other parameters are fixed at $K_{2}=8$, $p=1$. Solid cyan and solid black curves represent the movement of two saddle-node bifurcation points SN1 and SN2 respectively. Dotted black and dashed black curves respectively represent two pitchfork bifurcation points PB2 and PB1. The bifurcation curves separates the whole spaces in four regimes based on the synchronization states, marked on the diagram. $h$ promotes multistability and $\beta$ promotes bistability in the $K_1-h$ space.}
    \label{fig11}
\end{figure}

\begin{figure*}
    \centering
    \includegraphics[scale=0.35]{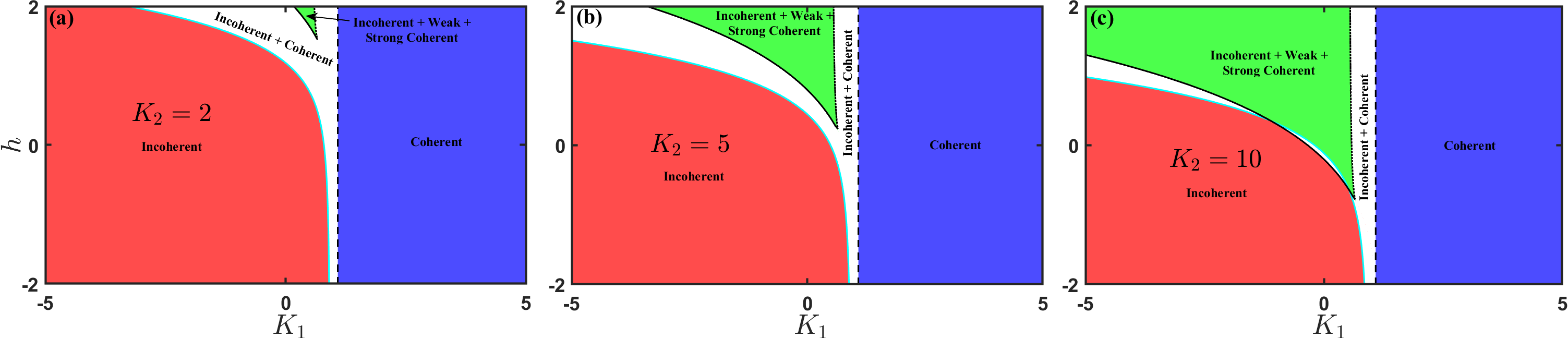}
    \caption{Stability diagram in $K_1-h$ parameter space for (a) $K_2=2$, (b) $K_2=5$ and (c) $K_2=10$ for $\beta=0.4$ and $p=1$ plotted using the reduced-order model. Solid cyan and solid black curves represent the movement of two saddle-node bifurcation points SN1 and SN2 respectively. Dotted black and dashed black curves respectively represent two pitchfork bifurcation points PB2 and PB1. The bifurcation curves separates the whole spaces in four regimes based on the synchronization states, marked on the diagram. Clearly, $K_2$ promotes multistability in the $K_1-h$ space.} 
    \label{fig10}
\end{figure*}

In Fig.~\ref{fig7} we have checked the effect of $K_2$ on the $K_1-p$ space by taking $K_2=2,~5$ and $8$. Here, also the bifurcation curves separate the whole regime into five parts. Starting from $K_2=2$, the red regime of the incoherent state, enclosed by the SN1 points, is wider and the green regime of multistable states are narrower than other regimes (white, yellow and blue). As we have increased the $K_2$ value, the curve of SN1 and SN2 points moves in the backward direction (negative $K_1$), implies the stability of the strong and weak synchronized state in repulsive pairwise coupling. Further increase in $K_2$ keeps the system synchronized in more negative coupling, indicates an increase in hysteresis width. These findings using the reduced order model are validated numerically in Fig.~\ref{fig8} by taking $p=-0.5$.

Next in Fig.~\ref{fig11} we have shown the effect of $\beta$ on $K_1-h$ space. Fig.~\ref{fig11}(a) illustrates the role of $h$ on the synchronization behavior of the considered system at a specified $\beta (=0.4)$ value. At lower $h$ value the parameter space is separated by the SN1 and PB1 curves, implying a classical synchronization transition. Naturally the system shows three states, incoherent, coherent and bistable. After a certain $h$, two curves indicating SN2 and PB2, appear in the parameter space. As a consequence, a weak synchronization state helps the synchronization transition in the green regime. The increase in $h$ broadened this regime for multistability. Despite the fact that increasing $h$ reduces the higher-order coupling strength, the hysteresis width increases. This is a counter intuitive result, because the higher-order coupling strength $K_2$ generally promotes bistability in a system. Therefore, the multilayer structure of the networked system is responsible for this behavior. Fig.~\ref{fig11}(b) reports that increase in $\beta$ changes the position of the bifurcation points, i.e the critical coupling strengths. Also, the white regime of bistability expanded and the green regime of multistability shrinks. Thus the phase-lag parameter $\beta$ affected the weak synchronization state for particular $h$ values.            

Completely opposite scenario has been observed when we have increased the value of $K_2$ by fixing other parameters in Fig.~\ref{fig10}. The higher-order coupling strength $K_2$ promotes multistability without affecting the onset of synchronization. The green regime expanded in the downwards by reducing the area of other three regimes. Interestingly, the pitchfork bifurcation points PB2 moves along the particular line $K_1=1.0857$. Furthermore, in Fig.~\ref{fig10}(c), the two saddle-node bifurcation curves (SN1 and SN2) intersect at two distinct points, located at $(K_1,h)=(0.5271,-0.6428)$ and $(K_1,h)=(-1.03,0.3263)$. For $h \geq 0.3263$ the difference between the SN1 and SN2 curves ensures the bistability between the incoherent and strongly synchronized state.

\begin{figure*}
    \centering
    \includegraphics[scale=0.35]{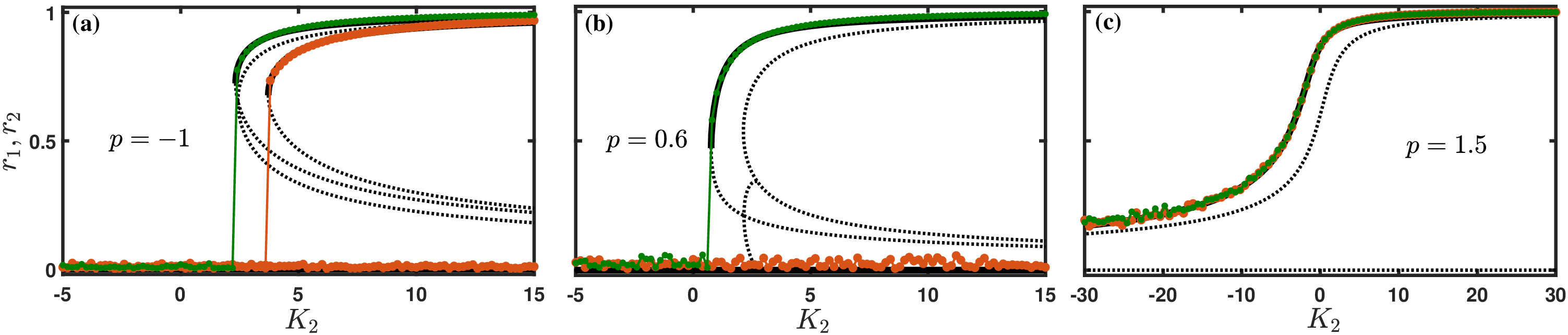}
    \caption{Synchronization profile with the variation of higher-order coupling $K_2$ for (a) $p=-1$, (b) $p=0.6$ and (c) $p=1.5$. Other parameters are fixed at $K_1=1.2$, $h=1$, and $\beta=0.4$. Black solid (stable) and dotted (unstable) $r_1,~r_2$ curves are plotted by solving the reduced order model. Green and brown filled circles joined by solid, dotted and dashed dotted lines correspond to the numerically simulated order parameter values from system (\ref{eqn1}). The fixed points changes stability through saddle-node bifurcations. $p$ promotes continuous transition to synchronization.}
    \label{fig14}
\end{figure*}

\begin{figure}
    \centering
    \includegraphics[scale=0.54]{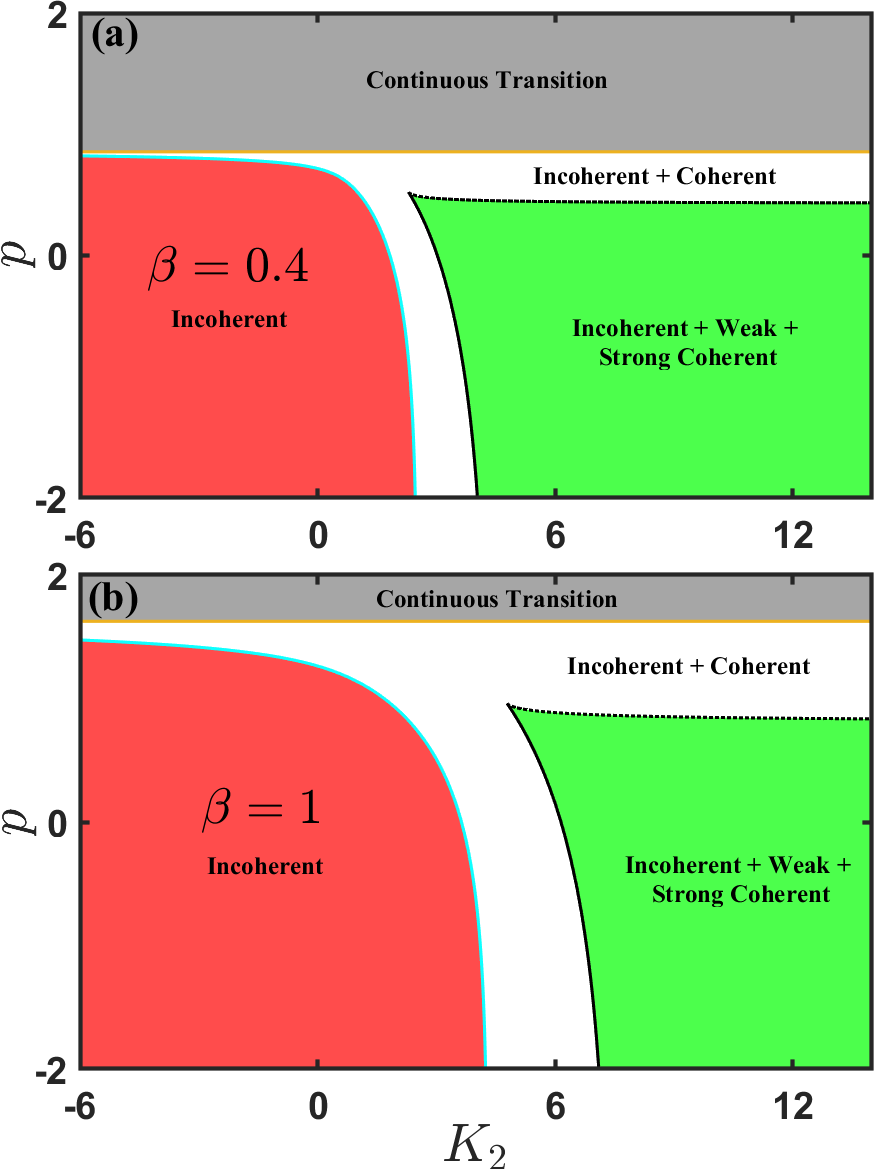}
    \caption{Stability diagram in $K_2-p$ parameter space for (a) $\beta=0.4$ and (b) $\beta=1$ plotted using the reduced-order model. Other parameters are fixed at $K_1=1.2$ and $h=1$. Solid cyan and solid black curves represent the movement of two saddle-node bifurcation points SN1 and SN2 respectively. Dotted black curve respectively represents the pitchfork bifurcation points PB2 and the solid yellow line represents the critical $p$ above which the transition becomes continuous. The bifurcation curves separates the whole spaces in four regimes based on the synchronization states, marked on the diagram.} 
    \label{fig12}
\end{figure}

So far we have checked the effect of the phase-lag parameter $\beta$ and higher-order coupling $K_2$ on the parameter spaces $K_1-p$ and $K_1-h$, provides a clear understanding of the role of the parameters on the synchronization behavior whenever $K_1$ is varied. Along with that, numerical validation of the analytical findings reconfirms the results. Now, we proceed to investigate the combined effect of the parameters $p,~h$ and $\beta$ when $K_2$ is varied. In Fig.~\ref{fig14} we plot the synchronization diagrams by increasing the value of $p$ from negative to positive. For $p=-1$ the reduced system go through only two saddle-node bifurcation points, leading the system to follow explosive synchronization transition. The saddle-node bifurcation points vanish one by one with increasing $p$. Therefore, for high $p$ the transition becomes continuous. Numerical data points agree well with the analytical ones. To better understand the effect of $p$ we have prepared $K_2-p$ space by tracing the movements of the bifurcation points and marked the regimes based on their synchronization states in Fig.~\ref{fig12}. Note that in Fig.~\ref{fig4}(b), $\beta$ is seen to promote multistability with the variation of $K_2$. To observe the role of $\beta$ on $K_2-p$ space we have plotted $K_2-p$ space for two different $\beta$. For $\beta=0.4$ the $K_2-p$ space is separated by the bifurcation curves into mainly two parts; one corresponds to the continuous transitions (gray regime) and another corresponds to the discontinuous transitions. The regime of discontinuous transitions is again separated in three parts, incoherent, bistable and multistable. Unlike Fig.~\ref{fig4}(b), here also the regime of continuous and discontinuous transition is separated by a critical parameter value. At low $p$ values, the fixed points of the reduced system changed their stability through two saddle-node bifurcation points. As the value of $p$ increases, the reduced system experiences a pitchfork bifurcation. For high $p$ values the transition becomes continuous. Now, we increase the $\beta$ value by keeping all other parameters fixed at the same value in Fig.~\ref{fig12}. We observed that $\beta$ expanded the bistable and multistable regimes and shrinks the regime for continuous transition.  

\begin{figure*}
    \centering
    \includegraphics[scale=0.35]{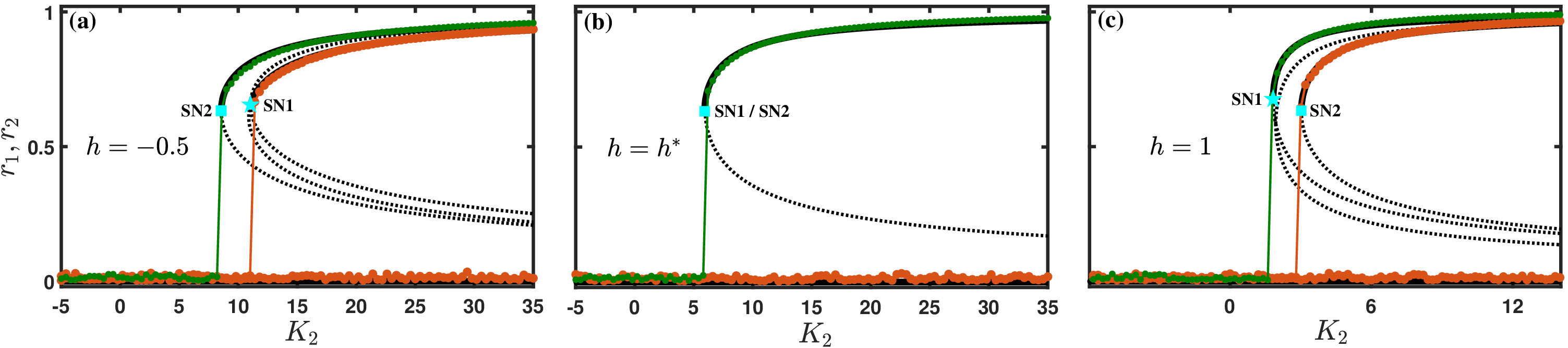}
    \caption{Synchronization profile with the variation of higher-order coupling $K_2$ for (a) $h=-0.5$, (b) $h=h^*=0.0182592$ and (c) $h=1$. Other parameters are fixed at $K_1=1.2$, $p=0$, and $\beta=0.4$. Black solid (stable) and dotted (unstable) $r_1,~r_2$ curves are plotted by solving the reduced order model. Green and brown filled circles joined by solid, dotted and dashed dotted lines correspond to the numerically simulated order parameter values from system (\ref{eqn1}). The fixed points change stability through saddle-node bifurcations. Clearly $h$ promotes bistability.}
    \label{fig16}
\end{figure*}

\begin{figure}
    \centering
    \includegraphics[scale=0.54]{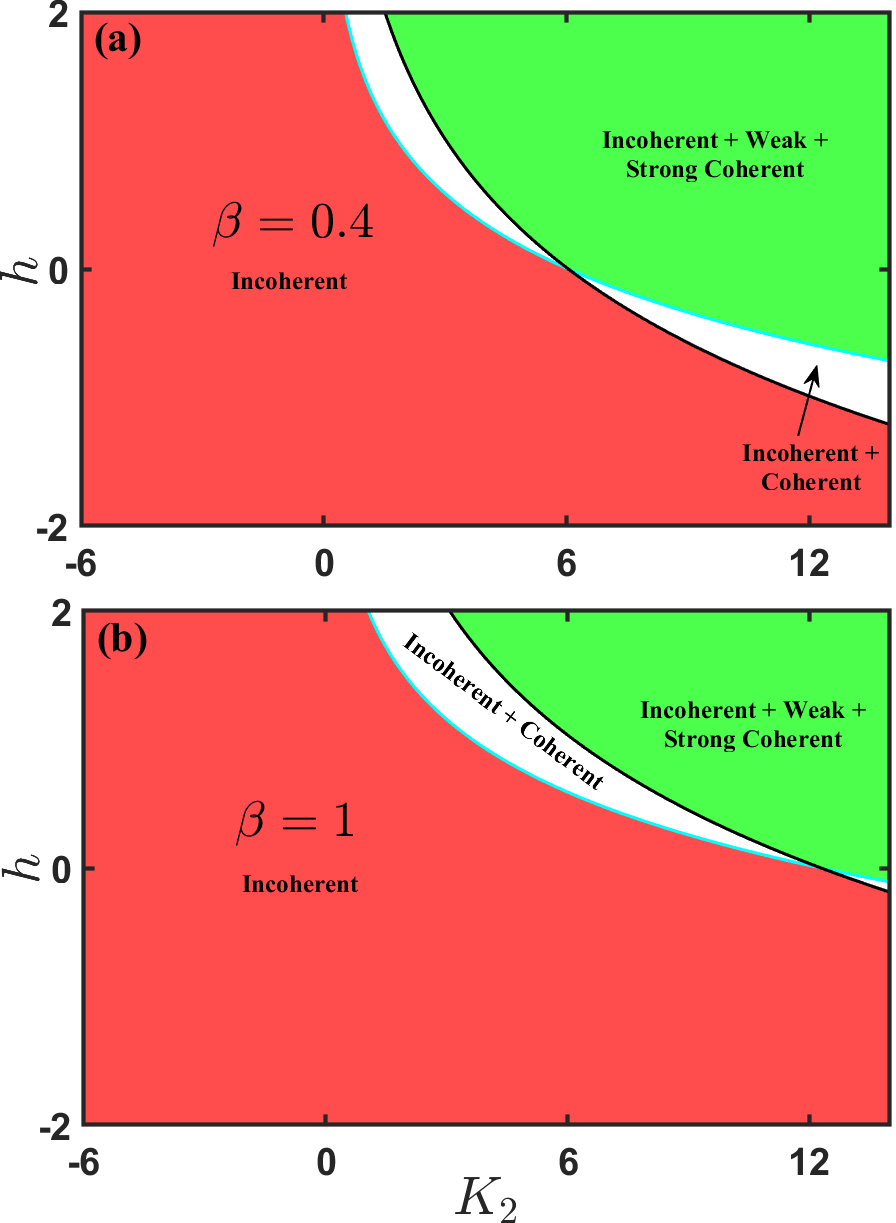}
    \caption{Stability diagram in $K_2-h$ parameter space for (a) $\beta=0.4$ and (b) $\beta=1$ plotted using the reduced-order model. Other parameters are fixed at $K_1=1.2$ and $p=0$. Solid cyan and solid black curves represent the movement of two saddle-node bifurcation points SN1 and SN2 respectively. The bifurcation curves separates the whole spaces in three regimes based on the synchronization states, marked on the diagram. $h$ promotes both bistability and multistability in $K_1-h$ space.}
    \label{fig15}
\end{figure}

In a similar manner, we have checked the role of $h$ on synchronization transition with variation of $K_2$. In Fig.~\ref{fig16} we present the synchronization diagrams for $h=-0.5,~0.018$ and $1$. For $h=-0.5$, the system shows explosive synchronization transition involving weak and strong synchronization state. At a critical $h$ two saddle node bifurcation points merge and the branchs for weak and strong synchronization state are also merge. For higher $h$ the reduced system again experiences two saddle-node bifurcation points, leading explosive synchronization with weak and strong synchronization state. To understand this behavior we have plotted $K_2-h$ space for two different $\beta$ in Fig.~\ref{fig15}. For $\beta=0.4$ two saddle node bifurcation curves divided the $K_2-h$ space into three regimes, incoherent, bistable and multistable. These two saddle node bifurcation curves intersect at $(K_2,h)=(5.97,0.018)$. As $h$ increases the area of the bistable states shrinks and vanishes at the intersection point of two saddle node bifurcation curves. The system again shows bistability for $h>0.018$. Fig.~\ref{fig15} demonstrates that higher $\beta$ value shifts the saddle-node bifurcation points to the higher $K_2$ value for each $h$. Also the two saddle-node curves intersect in high $K_2$. As a result the regime of incoherent state broadened and the regime of the multistable state decreases.     


\section{Conclusion}
In summary, this work presents the role of the phase-lag parameter and the higher-order (triangular) coupling strength on synchronization behavior of an adapted multilayer (two-layer) network. Here the nodes of one layer is interacting to the nodes of the other layer through the global order parameter adaptation, i.e. the order parameter corresponding to one layer has been adapted to another layer through some power function. Using the Ott-Antonsen ansatz we have reduced the dimension of the considered system in the thermodynamic limit. This allows to investigate the synchronization transitions of the system at low computational cost. Also the agreement of numerically simulated order parameter values of the considered $N-$dimensional system with the analytically obtained order parameter values from the reduced order model at some points of the parameter spaces validates the analytical findings. Firstly we observed that unlike~\cite{dutta2023impact, dutta2024transition} here also the phase-lag parameter $\beta$ plays dual role with the variation of pairwise and higher-order interactions. When the order parameters $r_1,~r_2$ are plotted as a function of $K_1$, the system exhibits both explosive and tiered synchronization transitions with proper choice of initial condition for $\beta=0$. The occurrence of two saddle-node and subcritical pitchfork bifurcation is the reason behind these two synchronization transitions. The increase in phase-lag value vanishes the weak synchronization state and the transition remains only explosive. On the other hand the phase-lag parameter induces multistabilty in both layers of the system with the variation of the higher-order coupling strength. The stability diagrams prepared using the reduced order model illustrate the impact of phase-lag.     

Along with that, we have studied the combined role of the parameters on the synchronization transitions by plotting the stability diagrams on the parameter spaces $K_1-p$, $K_1-h$, $K_2-p$ and $K_2-h$. Two interesting synchronization behavior have been found. One is although the increase in the exponent $p$ decreases the effective pairwise coupling strength $K_1$, it reduces the hysteresis width of the transition (with $K_1$ variation) in spite of the dominance of $K_2$ in the system. On the other hand the exponent $h$ promotes multistability in spite of reducing the effective higher-order coupling strength. Similar kind of things happened with the variation of $K_2$. Therefore, due to the multilayer configuration of the network and order parameter adaptation the system shows this counterintuitive behavior. This framework can be generalized to study the synchronization behavior of systems with multiple layers having more than triangular interactions.  
\section{ACKNOWLEDGMENT}
S.D. acknowledges the support from DST, India under the INSPIRE program (Code No. IF190605).

\newpage

%

\end{document}